\documentstyle[11pt]{article}                                                  
\newcommand{\beq}{\begin{equation}}
\newcommand{\eeq}{\end{equation}}
\newcommand{\bea}{\begin{eqnarray}}
\newcommand{\eea}{\end{eqnarray}}
\newcommand{\bal}{\begin{array}{ll}} \newcommand{\eal}{\end{array}}

\def\tag#1{\space}%
\def\text#1{\space}%
\def\stackunder#1#2{\mathrel{\mathop{#2}\limits_{#1}}}%
\def\bi{\bar{\imath }}
\def\bj{\bar{\jmath }}
\def\hi{{\imath}}
\def\hj{{\jmath}}
\def\hk{ k}
\def\ha{ a}
\def\hc{ c}

\def\hbj{{\bar{\jmath}}}
\def\hbk{{\bar k}}

\def\hbc{{\bar c}}
\def\bal{\bar{\alpha}}
\def\bbe{\bar{\beta}}
\def\ve{\varepsilon_{1}}
\def\vbe{{\bar {\varepsilon}}_{1}}
\def\vee{\varepsilon_{2}}
\def\vbee{{\bar {\varepsilon}}_{2}}
\def\lam{\lambda}
\def\diff#1#2{{{\partial #1}\over{\partial #2 }}}

\begin{document}
\begin{titlepage}

\begin{flushright}
Saclay T96/065 \\
hep-ph/9606383 \\
June 1996\\
\end{flushright}

\vskip.5cm
\begin{center}
{\huge{\bf Abelian Flavour Symmetries in Supersymmetric Models \footnote{Supported in part by the EC grant Flavourdynamics.}}}
\end{center}
\vskip1.5cm

\centerline{ E. Dudas $^a$, C. Grojean $^{a,b}$, S. Pokorski 
\footnote{Supported in part by the Polish 
Committee for Scientific Research. } $^c$ and C.A. Savoy $^a$}
\vskip 15pt
\centerline{$^{a}$ CEA-SACLAY, Service de Physique Th\'eorique}
\centerline{F-91191 Gif-sur-Yvette Cedex, FRANCE}
\vskip 3pt
\centerline{$^{b}$ Ecole Normale Sup\'erieure de Lyon}
\centerline{F-69364 Lyon Cedex 07, FRANCE}
\vskip 3pt
\centerline{$^{c}$ Institute of Physics, Warsaw}
\vglue .5truecm
\begin{abstract}
\vskip 3pt
We propose a theory of flavour based on abelian horizontal
gauge symmetries and modular invariances.  We construct explicit
supergravity models where the scale of the horizontal $U(1)$ 
symmetry breaking is fixed by the Green-Schwarz mechanism for
anomaly cancellation. The supersymmetric spectrum is obtained in terms 
of the $U(1)$ charges which are determined by the Yukawa matrices.
\end{abstract}

\end{titlepage} 
\section{Introduction}

There is a revival of interest in
explaining the pattern of fermion masses and mixings by postulating a
horizontal $U(1)_X$ gauge symmetry \cite{ibanez94}-\cite{nir95}.
 The $U(1)_X$ charges are assigned to fermions in such a way that only
a small number of Yukawa interactions is allowed by the symmetry. The
remaining effective Yukawa vertices are generated through non-renormalizable
couplings to fields which are Standard Model gauge singlets but carry 
horizontal charge, in the effective theory defined at some large scale $M_P$
(we shall identify it later with the Planck scale).
When these singlet fields get vev's of $O(\varepsilon M_P ),\varepsilon \le 1$ and 
spontaneously break $U(1)_X$, the resulting Yukawa couplings are
suppressed by powers
of the small parameter, $\varepsilon ^{n_i}$,  where the powers $n_i$ 
depend on the $U(1)_X$ charge assignment \cite{froggatt}. There are several reasons to assume this 
horizontal symmetry to be a local one. Of course, this avoids physical
problems related to massless Goldstone bosons. Moreover, in the
context of supersymmetric models with ``stringy'' $U(1)_X$ symmetry 
\cite{dine87}-\cite{font88}, this mechanism of fermion mass generation shows an
interesting connection between phenomenologically viable mass pattern and
the Green-Schwarz mechanism of anomaly cancellation, which successfully
predicts the Weinberg angle \cite{ibanez93}. It has also been suggested
that the Fayet-Iliopoulos term that is fixed by the anomalies will naturally
generate the small parameter $\varepsilon $ in the $U(1)_X$ breaking.

 In broken supergravity models with horizontal abelian symmetries the 
squark soft masses and the trilinear soft terms are correlated 
with the quark mass matrices by the symmetry. 
Generically, the off-diagonal entries (in the basis in which quarks are 
diagonal) are predicted \cite{leurer93} to be of the order of
some powers of the Cabibbo mixing angle $\lambda$ and in some cases
comparable with the existing experimental limits. In a recent paper
\cite{DPS96} it has been proposed to impose on such models the symmetries
(and the spectrum) generic for effective supergravity lagrangian which
originates from orbifold models of string compactifications. 
By combining the horizontal symmetry and the modular invariances that
characterize these models, the soft terms
can be calculated not only as powers of $\lambda$ but also with their
relative coefficients fixed in terms of the horizontal charges and
modular weights. 

If one consistently assumes that the hierarchies in the fermion 
mass spectrum are entirely due to the $U(1)_X$ symmetry, there are 
interesting relations between charges and modular weights. As a 
consequence, the sfermion (squarks and sleptons) spectrum is 
also predicted in terms of the $U(1)_X$ charges. 
Given a pattern of fermion mass matrices in terms of a 
set of $U(1)_X$ charges, the sfermion masses are obtained in terms of two
parameters characterizing the supersymmetry breaking, the gaugino mass
$ M $ and the gravitino mass $m_{3/2} .$ Now, this provides a unique 
way to test our horizontal symmetry models by a study of the sfermion
mass spectrum, since the fermion masses and mixings are used to fix
the charges and the $U(1)_X$ sector is in general too heavy to become
visible.    

By adopting the dilaton/moduli parametrisation of supersymmetry 
brea\-king proposed in \cite{brignole1} and in the case of one
$U(1)_{X}$ symmetry spontaneously broken by the vev of a scalar field of
charge $X=-1$, we get \cite{DPS96} (in the $U(1)_X$ basis)
\bea
\widetilde{m}_{q_i}^2 - \widetilde{m}_{q_j}^2 
& = & 
(q_i - q_j) \ m_{3/2}^2 \ , \nonumber \\  
\widetilde{m}_{q_i}^2 +
\widetilde{m}_{u_j}^2 +
\widetilde{m}_{h_2}^2
& = & 
M^2 + (q_i + u_j + h_2) \ m_{3/2}^2 \ , 
\label{1}
\eea
where  $\widetilde{m}_{q_i}$ ($\widetilde{m}_{u_i})$ is the diagonal element of
the sfermion mass 
matrix and $q_i$ ($u_i$) the $U(1)_X$ charge associated with the left-handed 
(right-handed up)  quark
of the $i^{th}$ family, and $\widetilde{m}_{h_2}$ and $h_2$ are the 
corresponding quantities for one of the Higgs doublets. 
Regarding the trilinear soft terms we get, e.g.,  the prediction
\bea
 A^{U}_{ij} & \simeq &
- M + (q_i + u_j + h_2) \  m_{3/2} \, 
\label{3} 
\eea
for the coefficient of trilinear coupling $Y_{ij}^U Q^i U^j H_2$.
The complete relations exhibit flavour off-diagonal terms in the soft 
masses and the full trilinear
soft terms are not proportional to the corresponding Yukawa
couplings. As will be explained, these additional
contributions give physical observable effects of the same
order as the "diagonal" ones displayed in (\ref{1}) and (\ref{3}).
Similar relations hold for down-type
squarks and sleptons.  It must be noticed that the simple results 
(\ref{1}) - (\ref{3}) follow from a cancellation between the geometrical
supergravity contribution and the part of the mass splitting from the $U(1)_X$
D-term that contain model dependent parameters. The horizontal splitting 
in  (\ref{1}) is proportional to the charge differences which
appear in the Yukawa mass matrices. Since the lighter fermions are associated
to larger charges, the corresponding sfermions are predicted to be
heavier than superpartners of heavier fermions.   

Extensions of the Standard Model of the electroweak interactions are constrained by flavour changing neutral current (FCNC) processes like 
$K^0-{\bar K}^0$ mixing, $b \rightarrow s \gamma$, $\mu \rightarrow e \gamma$ 
or the electric dipole moment (e.d.m.) of the neutron. The observed 
suppression of FCNC
transitions is nicely explained in the Standard Model (SM) by the GIM
me\-chanism.  The supersymmetric extensions of the SM do contain
additional contributions to FCNC transitions from sfermion exchange in loop
diagrams. They can be potentially dangerous if, in the basis 
in which fermion mass matrices are  diagonal,
the sfermion mass matrices have large flavour off-diagonal entries.
In addition, new phases which are usually present in the sfermion mass matrices
and in the trilinear couplings can give too large CP violating effects
(e.g. too large neutron e.d.m.).
After a rotation to the quark diagonal basis, we obtain our prediction
for the squark mass matrices up to an overall scale. 
The question one may ask in our class of models is this: 
can one suppress the off-diagonal entries in
the squark mass matrices below the order of magnitude estimate based on the
$U(1)_X$ symmetry alone? A priori, this might occur if some coefficients
vanish, i.e. for certain choice of horizontal charges. The conclusion of
ref.\cite{DPS96} is that such an additional off-diagonal suppression does not 
hold for the phenomenologically acceptable quark masses. 
In the case of one $U(1)_X$ symmetry,
squarks masses can be neither universal nor aligned with quark masses.Thus, the 
models of this class predict some deviations
from the Standard Model predictions for the FCNC transitions to be eventually
observed at higher level than in the case of universal soft masses at the
Planck scale. 

In the next three  Sections  of this paper we discuss the case of one $U(1)_X$
symmetry. We extend the formalism of ref.\cite{DPS96} by allowing for
flavour dependent modular weights for the superpotential and give some details
of the calculation of the soft terms. The breaking of the horizontal 
$U(1)_X$ is investigated and the induced D-term is
evaluated. In Section  \ref{sec:Higgs} , the Higgs sector mass parameters
are obtained in terms of their $U(1)_X$ charges, which are restricted by
the fermion masses. Then, it is shown that the
requirement of proper electroweak symmetry breaking strongly constrains the
scales $m_{3/2}$ and $M$. In consequence, the magnitude of the FCNC effects
is also determined. It is interesting to notice that this highly predictive
class of models with one $U(1)_X$ symmetry, although only marginally acceptable
from the point of view of FCNC effects, is qualitatively consistent with the
existing phenomenology and, moreover, gives testable predictions for
superpartners masses.

In Section \ref{sec:2U1} we consider a class of models with two abelian horizontal symmetries. They have been suggested \cite{leurer93} to reduce
flavour non-diagonal entries in the scalar mass matrices and so to alleviate
FCNC effects. By an ad hoc choice of charges, one can further suppress
$\bar{K} K$ and $\bar{D} D$ mixings. 
The relation between the modular weights and horizontal
charges are derived for this case and used to calculate the soft mass
terms. It becomes clear from our results that, with any number of
abelian symmetries, it is impossible to get either universal or
aligned (with quarks) squark masses for acceptable
quark mass matrices and therefore the models predict interesting phenomenology
in the FCNC sector.   

In the last section we summarize the main features
of the models with horizontal $U(1)'s$ and modular invariances from the
phenomenological point of view.


\section{Yukawa matrices from horizontal $U(1)_X$ gauge symmetry} \label{sec:generalites}

A natural way of understanding the fermion mass matrices is to postulate a family 
(horizontal) gauge symmetry spontaneously broken by the vacuum expectation
values (vev's) of some scalar fields $\phi$ which are singlets under the
Standard Model  gauge group. The hierarchy of fermion masses and mixing
angles is then explained by assigning different charges to different
fermions.  Only the third family of fermions acquires a mass at the tree
level and all the
other Yukawa couplings are forbidden by the $U(1)_X$ symmetry. After 
spontaneous symmetry breaking of the $U(1)_X$ symmetry, higher order invariant
terms in
the lagrangian (or superpotential in the
supersymmetric case) can be written and have the form 
$({<\phi> \over M_P})^{n_{ij}} \bar \psi_i
\psi_j H$ (after decoupling of the heavy fields), where $\psi_i$ are the
SM fermions, $H$  is a Higgs field and $M_P$ is a large scale.
Postulating $ \varepsilon  \equiv  {\langle
\phi\rangle \over M_P} \simeq \lambda$ (the Cabibbo angle) one can easily
explain hierarchies in the effective Yukawa couplings, precisely in the 
simplest case of abelian symmetry, with all the coefficients of the
higher dimension operators of the order $O(1)$.

In the context of the string inspired models, the most natural candidate
for such a symmetry is the anomalous $U(1)_X$ gauge group present in most
of the known $4$-dimensional string models \cite{dine87,font88}.
In this 
case, one-loop triangle graphs generate Adler-Bell-Jackiw gauge
anomalies. 
The gauge symmetry is restored by the $4$-dimensional version of the Green-
Schwarz mechanism \cite{GS}, by the use of the dilaton-axion superfield.
The relevant terms in the Lagrangian are
\bea
{\cal L}_{gauge} = {1 \over 4} \sum_a k_a S \int d^2 \theta ( \ tr W^{\alpha}
W_{\alpha})_a \ , \label{05}\\
K = - ln (S + S^+ + \delta_{GS} V) + ... \ . \nonumber
\eea
In (\ref{05}), $V$ and $W_{\alpha}$ are respectively the $U(1)_X$ gauge superfield and the
gauge superfield strenght, $S$ is the dilaton superfield,
$k_a$ is the Kac-Moody level of the factor $G_a$ of the
gauge group 
$G=SU(3)_C$$\bigotimes SU(2)_L$$\bigotimes U(1)_Y$$\bigotimes U(1)_X,$
and $\delta_{GS}$ is the Green-Schwarz coefficient. 
Under a $U(1)_X$ gauge transformation, $S$ is shifted as 
$S \rightarrow S + i \delta_{GS} \alpha (x)$. The complete
Lagrangian is gauge invariant provided the anomaly coefficients $A_i$ satisfy 
the condition
\bea
\delta_{GS} = {A_1 \over k_1} = {A_2 \over k_2} = {A_3 \over k_3} =
{A_X \over k_X} \ , \label{06}
\eea
where $\delta_{GS}$ is computed to be 
$\delta_{GS} = {1 \over 192 \pi^2} \ Tr X$. The mixing $S-V$ in the K\"ahler
potential gives rise to a term ($g_X = 1/(S+S^{\dagger})$)
\bea
V_D = {g_X^2 \over 2} \left( \sum_A K_A X_A \phi^A + {M_P^2 \over 192
\pi^2} \ Tr X \right)^2
 \ , \label{060}
\eea
in the scalar potential, where $A$ labels all chiral
fields of charges $X_A$ and $K_A = {\partial K \over \partial \phi^A}$.
The last term in the parenthesis in (\ref{060})
is the coefficient of the Fayet-Iliopoulos D-term 
\cite{FI}, which forces at least one of the fields to get a vacuum
expectation value and to break the $U(1)_X$ gauge symmetry 
at a scale slightly
below the Planck scale, depending on the value of $TrX$. 
If $n>1$ $U(1)$ symmetries are
considered, 
we can always define $n-1$ anomaly free symmetries and the present 
discussion applies.  Recently, it has been  shown \cite{ibanez94} that  
using this symmetry and imposing phenomenologically successful fermion mass
matrices one correctly predicts the Weinberg angle 
at the scale where $U(1)_X$ is spontaneously broken. 
This indicates a close relation between the
Green-Schwarz mechanism and $U(1)_X$ fermion mass matrices. Also,
a direct connection between fermion mass matrices and mixed gauge
anomalies was established \cite{binetruy1} and further studied in 
\cite{dudas95,nir95}. Within this scheme
the scale $M_P$ of the effective theory
is assumed to be the Planck mass, while 
the parameter $ \varepsilon^2 $ is equal to 
$ -(M_P^2 Tr X )/ (192\pi^2 X_\phi) $.   
 
Let us now consider the case when one SM gauge singlet takes a vev 
and all other matter field vev's are zero at the Planck scale.
Supersymmetry is assumed, so that the 
Yukawa couplings are encoded in the $U(1)_X$ invariant superpotential 
\begin{eqnarray}
W  &=&
\sum_{ij}
\left[ Y_{ij}^U\;
\theta \left( q_i+u_j+h_2\right) 
\left( \frac \phi {M_P}\right)
^{q_i+u_j+h_2}Q^iU^jH_2\right. 
\nonumber \\
&&
\hskip .8cm
+Y_{ij}^D\;
\theta \left( q_i+d_j+h_1\right) 
\left( \frac \phi {M_P}\right) 
^{q_i+d_j+h_1}Q^iD^iH_1  
\nonumber \\
&&
\hskip .8cm
+\left. Y_{ij}^E\;
\theta \left( \ell _i+e_j+h_1\right) 
\left( \frac \phi {M_P}\right) 
^{\ell _i+e_j+h_1}L^iE^jH_1\right] , 
\label{superpo}
\end{eqnarray}
\noindent where we denote the matter fields by
capitals $\Phi ^i$, 
the corresponding $U(1)_X$ charges by small letters 
$\varphi _i$ (after choosing the normalization of the  $U(1)_X$ to be
such that the singlet charge is $X_\phi = -1.$)
The $\theta$-functions remove the Yukawa couplings that are forbidden by the $U(1)_X$ symmetry combined with the holomorphicity of $W$. 
We also assume R-parity symmetry in $W$ (see \cite{BLR} for attempts to
enforce R-parity through horizontal symmetries). All the allowed entries in the 
Yukawa coupling matrices $Y^D, \, Y^D, \, Y^E$ are assumed to be "natural", 
of $O(1)$. The scalar potential in 
(\ref{060}) vanishes for  
\bea
\phi^2 = \epsilon^2 M_{P}^2 = {{TrX} \over {192 \pi^2}} M_P^2 \ ,
\label{phi}
\eea
if we postulate $TrX\, > \, 0$. As stated before we assume $\epsilon$ to be of the order of the Cabibbo angle. By a choice of the $U(1)_X$ charges 
we get the powers of $\epsilon$ in the effective low energy Yukawa couplings,
\bea
{\hat Y}^U_{ij} \, = \, Y^U_{ij} \epsilon^{(q_i+u_j+h_2)} \theta (q_i+u_j+h_2)
\label{Yuka}
\eea
(analogously for ${\hat Y}^D, \,{\hat Y}^E$) needed to implement their hierarchy.

Before proceeding to study the consequences of this broken horizontal symmetry in 
the supergravity framework, let us pay some attention to the question of the uniqueness of the solution (\ref{phi}). 
Indeed, in the supersymmetric limit one expects degenerate minima of the scalar potential. 
They correspond to the vanishing of all the auxiliary fields: the gradients of 
the superpotential (F-type) and the D-terms associated to each one of the 
$SU(3)_{C}$$\bigotimes SU(2)_{L}$$\bigotimes U(1)_{Y}$$\bigotimes U(1)_X$ 
gauged symmetries. It has been shown that the D-terms vanish at and only at
those values of the field that correspond to extrema of holomorphic invariant polynomials \cite{group}
(with the exception of D$_X$-terms that contain a Fayet-Iliopoulos constant) .
This allows for a systematic classification of the zeros of the $SU(3)_{C}$$\bigotimes SU(2)_{L}$$\bigotimes U(1)_{Y}$ D-terms. 
For instance, the invariant $Q^i D^j H_1$ corresponds to a direction $Q^i = 
D^j = H_1 = v$ in the field space where all these D-terms vanish, while $D_X =
 0 $ may be written as%
\footnote{The vev's considered here are to be understood in the canonical basis as discussed in the next sections.}
\bea
g_X \left( -|\phi|^2 + (q_i+d_j+h_1) v^2 + \epsilon^2 M_P^2 
\right) = 0 \ .
\label{dx=0}
\eea

Combining similar solutions corresponding to all other invariants,
one defines a  manifold of solutions of D-terms. Those that corespond to
non-vanishing gradients of the superpotential $W$ break supersymmetry and are 
not minima of the scalar potential (at large scales). 
So, in our example (\ref{dx=0}), one easily checks from (\ref{superpo}) that for $v \neq 0$ and $\phi \neq 0$, the scalar potential will vanish only if all the 
entries in the i$^{th}$ row and j$^{th}$ column of the matrix $Y^D$ vanish. 
This is obviously excluded in phenomenologically viable models. Of course the 
same arguments applies for solutions associated to the $Y^U$ and $Y^E$ terms in 
(\ref{superpo}). 
The solutions of (\ref{dx=0}) with $\phi=0$ are also excluded if $q_i+d_j+h_1 \geq 0 $, $q_i+u_j+h_2 \geq 0 $  and  $l_i+e_j+h_1 \geq 0 $.
Therefore these conditions are sufficient ( though they seem also necessary 
at this stage, they can be avoided as discussed herebelow ) 
to ensure the uniqueness of the solution (\ref{phi}) if R-parity is assumed.
Next, one has to take care of solutions coming from R-parity violating invariants such as $L^k Q^i D^j$, 
where the Higgs $H_1$ is replaced in (\ref{dx=0}) by one of the sleptons $L^k$, 
and $L^k=Q^i=D^j \neq 0$. For $\phi \neq 0 $, the scalar potential gets 
a non-vanishing contribution if $Y^D_{ij} \neq 0$, ie if $q_i + d_j + h_1 \geq 0$. Analogously, from $L^kL^iE^j$ invariants one deduces the condition $h_1+l_i+e_j \geq 0$.
 Since the experimental data allow for some violation of R-parity, one can replace the positivity condition above by  a condition on the non-vanishing of rows and columns of the R-parity violating Yukawa couplings. 
In this case, the solutions of (\ref{dx=0}) with $\phi=0$, when allowed, will
be degenerate with that in (\ref{phi}). 
They correspond, however, to very isolated points in the field space and (\ref{phi}) should remain stable, if not unique.
As mentioned earlier, the positivity conditions are sufficient but not
necessary.
Other invariants with more than three matter and Higgs superfields can be included in the superpotential without any consequence on the Froggat-Nielsen mechanism. They may help to avoid the supersymmetric minima. For instance, a (R-parity conserving) term like $(H_1Q^iD^j)^2$ in $W$ would allow for a zero in the $Y^D$ matrix with $h_1+q_i+d_j < 0 $.
Notice that the (sufficient) positivity conditions are consistent with most of 
the models considered in the literature \cite{ibanez94}-\cite{nir95}.  

Finally, let us consider the two bilinear invariants $H_1H_2$ and $L^kH_2$. 
The vanishing of D-terms leads to $H_1=H_2=v$ and $-\phi^2 + (h_1+h_2)v^2 + \epsilon^2 M_P^2 = 0 $, for the former, and similarly for the latter with $H_1 \rightarrow L^k$ (the combination of these solutions are, in general, forbidden by non-zero entries in $Y^E$). 
However, at this level, one cannot argue that the degeneracy is lifted by  the
term $\mu H_1 H_2$ in the superpotential, since it has to be absent in the
supersymmetric limit, even if $(h_1+h_2) > 0$. For phenomenological reasons,
it has to be related to the supersymmetry breaking scale 
(the so-called $\mu-$problem). 
We postpone the question of this particular degeneracy until section $4$.

\section{$U(1)_X$ and modular symmetries in effective supergravity 
from string models} \label{sec:U(1)}

The phenomenologically interesting low energy limit of the superstring models
is
the $N=1$ supergravity defined by the K\"{a}hler function $K$, the
superpotential $W$ and the gauge kinetic function $f$.
The fields in the massless string spectrum are a universal dilaton $%
S $, moduli fields with no scalar potential generically denoted by $T_\alpha $ and matter chiral
fields $\Phi ^i$ which include the SM particles, 
$\Phi ^i=Q^i,U^i,D^i,L^i,E^i,H_1,H_2.$  An important role
in the following discussion will be played by the target-space modular
symmetries $SL(2,Z)$ \cite{kikkawa84} associated with the moduli
fields $T_\alpha (\alpha =1..p),$ acting as
$ T_\alpha  \rightarrow  (a_\alpha T_\alpha -i b_\alpha )$/$(ic_\alpha
T_\alpha +d_\alpha ),$ with $( a_\alpha d_\alpha -b_\alpha c_\alpha )=1 $
 and $ a_\alpha ...d_\alpha \in Z .$  In effective string theories of the
 orbifold type, the matter fields $\Phi
^i $ transform under $SL(2,Z)$ as
$\Phi ^i  \rightarrow ( ic_\alpha T_\alpha + d_\alpha )
^{n_i^{(\alpha )}}\Phi ^i $, where the $n_i^{(\alpha )}$ are called the modular weights 
of the fields $\Phi ^i$ with respect to the modulus $T_\alpha $ 
\cite{guetic91}.
These modular 
transformations, which are symmetries of the 
supergravity theory, can be viewed as a particular type
of K\"{a}hler transformations. 

We assume the existence of superstring models which in the low-energy limit
yield effective supergravity theories with the above minimal content of
superfields,
the gauge group $SU(3)_{C}$$\bigotimes SU(2)_{L}$$\bigotimes U(1)_{Y}$$\bigotimes U(1)_X$ and a SM singlet supermultiplet $\phi $ with 
$X_\phi=-1$.
The superpotential $W $  is defined in (\ref{superpo}) and the
 K\"{a}hler potential $K$ is as follows,
\begin{eqnarray}
K &=&
{K_0}
\left( T_\alpha ,\bar{T} ^{\bal} \right) 
-\ln \left( S+\bar{S} \right)  + 
\stackunder{\alpha =1}{\stackrel{p}{\Pi }}t_\alpha
^{n_{\phi }^{(\alpha )}} \bar{\phi } \phi 
\nonumber \\
&& \ \ + \sum_{\Phi ^i=Q^i,U^i,D^i,L^i,E^i,H_1,H_2} K^{\Phi}_{i \bj}
\Phi ^i {\bar{\Phi }}^{\bj }  ,  
\nonumber \\
&&  
\nonumber \\
K^{\Phi}_{i \bj} & = &
{\delta}_{i j}
\stackunder{\alpha =1}{\stackrel{p}{\Pi }}t_\alpha^{n_i^{(\alpha )}} + 
{Z}^{\Phi}_{\bi j}
\left[ 
\vphantom{\left( \frac {\bar{\phi}} {M_P}\right) ^{\varphi _j-\varphi _i}}
\,\,
\theta \left( \varphi _i-\varphi_j\right) 
\stackunder{\alpha =1}{\stackrel{p}{\Pi }}
t_\alpha^{n_j^{(\alpha)} + {\bar n}_{\Phi,i\bj}^{(\alpha)}}
\left( \frac{{\phi}}{M_P}\right) 
^{\varphi _i-\varphi _j}
+ \right.  
\nonumber \\
&&
\hskip 1.1cm
\left. 
\theta \left( \varphi _j-\varphi _i\right) 
\stackunder{\alpha =1}{\stackrel{p}{\Pi }}
t_\alpha ^{n_i^{(\alpha)} + n_{\Phi,i\bj}^{(\alpha)}}
\left( \frac {\bar{\phi}} {M_P}\right) 
^{\varphi _j-\varphi _i}
\,\,\right] + ....  \label{Kahler} 
\end{eqnarray}
where $i,j=1,2,3$. In (\ref{Kahler}), $t_\alpha $ are the
real part of the $p$ moduli fields $T_\alpha $ and the dots stand for higher
order terms in the fields $\phi$ and $\Phi ^i$.  Note that the flavour 
non-diagonal terms in the K\"{a}hler potential, proportional to the 
coefficients $Z^{\Phi}_{i \bj}$, are constrained only by the gauge symmetry and
R-parity and have the form
$K^{Q}_{i\bj}Q^i{\bar Q}^{\bj}$, 
$K^{U}_{i\bj}U^i{\bar U}^{\bj}$,
$K^{D}_{i\bj}D^i{\bar D}^{\bj}$,
$K^{L}_{i\bj}L^i{\bar L}^{\bj}$ and 
$K^{E}_{i\bj}E^i{\bar E}^{\bj}$.
The explicit dependence on $t_{\alpha}$ in these terms is fixed by the
modular invariance conditions, which are discussed in detail below.  
In general, the coefficients $Z^{\Phi}_{i \bj}$ are automorphic functions of
the moduli, of chiral weight
$n_{\Phi,i\bj}^{(\alpha)}$ and antichiral weight 
${\bar n}_{\Phi,i\bj}^{(\alpha)}$, i.e.
they transform under the modular transformations as 
$Z^{\Phi}_{i \bj} \rightarrow
(i c_{\alpha} T_{\alpha} + d_{\alpha})^{n_{\phi,i\bj}^{\alpha}}
(-i c_{\alpha} T_{\alpha}^+  + d_{\alpha})^{{\bar n}_{\phi,i\bj}^{\alpha}}
Z^{\Phi}_{i \bj}$. The coefficients
$Y_{ij}^U,Y_{ij}^D,Y_{ij}^E$ in (\ref{superpo})  can also be automorphic 
functions of the moduli fields, of weight $n_{U,ij}^{(\alpha)}$, etc.
Note that a coefficient with two analytic indices, like 
$n_{U,ij}^{(\alpha)}$, is related to the modular transformations of a Yukawa coefficient, here $Y^U_{ij}$,
while a coefficient with an analytic and an antianalytic indices, like 
$n_{U,i \bj}^{(\alpha)}$,  is related to the modular transformations of a non-diagonal K\"{a}hler coefficient, here 
$K^{U}_{i\bj}$. 

In order to impose the modular symmetries, let us first define 
$n_0^{(\alpha )}$ by the modular transformations of the  K\"{a}hler 
potential for the moduli fields, $K_0 \rightarrow K_0+
n_0^{(\alpha )} \ln \left| ic_\alpha T_\alpha +d_\alpha \right| ^2,$
which is a K\"{a}hler transformation. 
A typical example is
 \begin{eqnarray}
{K}_0 &=&- \sum_{\alpha =1}^p n_0^{(\alpha )} \ln \;t_\alpha .
\label{6} \end{eqnarray}
\noindent The modular invariance of the full K\"ahler potential 
requires the following
relations between modular weights and $U(1)_X$ charges 
\begin{eqnarray}
\left( \varphi_i-\varphi_j \right) n_\phi ^{(\alpha )} &=&
X_\phi \left(
n_i^{(\alpha )}-n_j^{(\alpha )} + n_{\Phi,i{\bj}}^{(\alpha )} -
{\bar n}_{\Phi,i{\bj}}^{(\alpha )}  \right) \, ,
\label{90} 
\end{eqnarray}
where $i,j = 1,2,3$ are family indices, $X_\phi$ is the $U(1)_X$ charge of the
singlet $\phi$ and $\varphi_i$ are $U(1)_X$ charges for fermions with the same SM quantum numbers.
So (\ref{90}) is a horizontal (family) relation applying separately for 
Q, U, D, L and E fermions.
  
From the superpotential, to be consistent with modular invariance of the 
complete theory, we get the following conditions 
for the quarks and leptons:
\begin{eqnarray}
-X_\phi^{-1} \left( q_i+u_j+h_2 \right) n_\phi ^{(\alpha )}+n_{q_i}^{(\alpha )}+n_{u_j}^{(\alpha
)}+n_{h_2}^{(\alpha )}+n_0^{(\alpha )}+n_{U,ij}^{(\alpha )} 
 &=& 
0 , 
\nonumber \\
-X_\phi^{-1} \left( q_i+d_j+h_1\right) n_\phi ^{(\alpha )}+n_{q_i}^{(\alpha )}+n_{d_j}^{(\alpha
)}+n_{h_1}^{(\alpha )}+n_0^{(\alpha )}+n_{D,ij}^{(\alpha )} 
&=& 
0,  \nonumber \\
-X_\phi^{-1} \left( l_i+e_j+h_1\right) n_\phi ^{(\alpha )}+n_{l_i}^{(\alpha )}+n_{e_j}^{(\alpha
)}+n_{h_1}^{(\alpha )}+n_0^{(\alpha )}+n_{E,ij}^{(\alpha )} 
&=& 
0.  
\label{8} 
\end{eqnarray}
\noindent Each of these equations must hold if the corresponding Yukawa term is not zero. From (\ref{8}), we get the relation
\begin{eqnarray}
\left( q_i-q_j \right) n_\phi ^{(\alpha )} &=&X_\phi \left(
n_{q_i}^{(\alpha )}-n_{q_j}^{(\alpha )} + n_{U,ik}^{(\alpha )} -
n_{U,jk}^{(\alpha )}  \right),  \label{9} 
\end{eqnarray}
\noindent as a consequence of the existence of the Yukawa couplings 
$Y_{ik}^U$ and $Y_{jk}^U$ in the superpotential.
Similar relations are obtained by replacing $q_i$ by $u_i,d_i,l_i,e_i$.
Comparing (\ref{9}) with (\ref{90}), we get additional conditions
between the modular transformations of the off-diagonal terms in the
K\"ahler potential and the transformations of the Yukawa couplings,
namely,
\begin{eqnarray}
&&n_{U,ik}^{(\alpha )}-n_{U,jk}^{(\alpha )}
=
n_{D,ik}^{(\alpha)}-n_{D,jk}^{(\alpha )}  
=
n_{Q,i{\bj}}^{(\alpha)}-{\bar n}_{Q,i{\bj}}^{(\alpha )}
\equiv s_{Q,ij}^{(\alpha)} \ , \nonumber \\
&&n_{U,ki}^{(\alpha )}-n_{U,kj}^{(\alpha )}
=
n_{U,i{\bj}}^{(\alpha)}-{\bar n}_{U,i{\bj}}^{(\alpha )}
\equiv s_{U,ij}^{(\alpha)} \ , \nonumber \\ 
&&n_{D,ki}^{(\alpha )}-n_{D,kj}^{(\alpha )}
= 
n_{D,i{\bj}}^{(\alpha)}-{\bar n}_{D,i{\bj}}^{(\alpha )}
\equiv s_{D,ij}^{(\alpha)}  \ , \nonumber \\
&&n_{E,ki}^{(\alpha )}-n_{E,kj}^{(\alpha )}
= 
n_{E,i{\bj}}^{(\alpha)}-{\bar n}_{E,i{\bj}}^{(\alpha )}
\equiv s_{E,ij}^{(\alpha)} \ , \nonumber \\
&&n_{E,ik}^{(\alpha )}-n_{E,jk}^{(\alpha )}
=
n_{L,i{\bj}}^{(\alpha)}-{\bar n}_{L,i{\bj}}^{(\alpha )}
\equiv s_{L,ij}^{(\alpha)} \ , 
\label{91}
\end{eqnarray}
which are all independent of the values of $k=1,2,3.$
Actually, by putting the $U(1)_X$ charges to zero in (\ref{90}) and (\ref{8}) 
one finds that the relations (\ref{91}) are independent of the
$U(1)_X$ symmetry and apply more generally.
They restrict the moduli dependence of the Yukawa couplings. 
Fourteen of these equations are independent, ten in the quark sector and four 
in the leptonic sector. So, if the quark Yukawa couplings are a priori defined
 by eigthteen modular weights, one for each coupling, only eight of them are
independent. Analogously, the lepton Yukawa couplings are defined by five
independent modular weights.

A simple and interesting case is when the Yukawa couplings do not depend on
the moduli fields or the dependence is flavour blind and Y are of $O(1)$.
We shall mainly consider this
case, to be consistent with our physical assumption in this paper,
that the fermion mass hierarchy is entirely due to
the $U(1)_X$ charges. Then the relations (\ref{91}) also
imply 
$n_{\Phi , i\bj}^{(\alpha)} = {\bar n}_{\Phi , i\bj}^{(\alpha)}$, etc., and
a real type modular transformations 
$Z^{\Phi }_{i \bj} \rightarrow
|i c_{\alpha} T_{\alpha} + d_{\alpha}|^{2 n_{\Phi ,i\bj}^{(\alpha)}}
Z^{\Phi }_{i \bj}$.
In this case (\ref{9})
simplifies to
\begin{eqnarray}
\left( \varphi_i-\varphi_j \right) n_\phi ^{(\alpha )} &=&X_\phi \left(
n_i^{(\alpha )}-n_j^{(\alpha )}  \right).  \label{92} 
\end{eqnarray}
The relations (\ref{92}) give a connection between the
modular weights and the $U(1)_X$ charges and
can be interpreted as an embedding of the $U(1)_X$ symmetry in the modular symmetries. They enable us to write the K\"{a}hler metric for the matter fields as
\bea
K^{\Phi}_{i \bj}&=&
\stackunder{\alpha =1}{\stackrel{p}{\Pi }}t_\alpha
^{(n_i^{(\alpha )}+n_j^{(\alpha )})/2} \left( {\delta}_{i \bj}+
Z^{\Phi}_{i \bj}{ \hat \varepsilon } ^ 
{| \varphi _i-\varphi _j|} \right)
\, ,
\eea
\noindent where the small
parameter ${\hat \varepsilon} = \prod_{\alpha} t_{\alpha}^{n_{
\phi}^{(\alpha)} / 2} {<\phi> \over M_P}$ settles the hierarchy in
the fermion and scalar mass matrix elements (if, for some $(i \bj),$ 
(\ref{92}) is not fulfilled, the coefficient vanishes).

If eq.(\ref{9}) are not satisfied, modular invariance of the superpotential
implies zeroes in the Yukawa matrices and
in the off-diagonal entries of the K\"{a}hler metric. These type of zeroes
must be distinguished from the ones given by $U(1)_X$ invariance and 
the holomorphicity of the
superpotential $W$ as described by the $\theta $-functions in (\ref{superpo}). We
could try to construct phenomenologically interesting models in this way, in
the spirit of ref.\cite{RRR}. However, as argued in \cite{DPS96}, due to
the fact that modular symmetry zeroes in Yukawa matrices imply zeroes
in the corresponding off-diagonal elements of the K\"{a}hler metrics, the
zero textures of the above matrices are preserved in the fermion canonical
basis. Phenomenologically, 
they can acommodate the fermion masses and one
mixing angle, but they cannot explain the whole $V_{CKM}$ matrix. Hence,
for the quarks, the
relations (\ref{9}) must be imposed for {\it all} the indices $(i,j)$ .

The physical Yukawa couplings $\hat Y$ are obtained by the canonical 
normalization of
the kinetic terms, which requires the redefinition of the fields
$\hat \Phi ^{\hi} = e^{\hi}_j \Phi ^j $ where the vielbein $ e^{\hi}_j
(t_\alpha , \phi )$
verifies
\begin{eqnarray}
K^{\Phi}_{i \bj}&=& \delta _{\hbk l} {\bar{e}}^{\hbk}_{\bj} e^{l}_i 
\, .
\label{12} 
\end{eqnarray}

The potential effect of these field redefinitions  
is to remove the eventual zeroes in the Yukawa matrices coming from the 
holomorphicity and $U(1)_X$ invariance of the superpotential
(Examples of this type of a phenomenological interest
can be found in \cite{dudas95}).

\section{Predictions for the soft terms } \label{sec:soft terms} 

The spontaneous breaking of local supersymmetry gives rise to a
low-energy global supersymmetric theory together with terms that 
explicitly break supersymmetry, but in a soft way. The signal of
supersymmetry breaking is in non zero vev's of the auxiliary
components of the chiral superfields $F^a=e^{\frac G2}G^a,$ where $G^a=
K^{a \bar{b} }\partial _{\bar{b} }G$ and $G = K + \ln |W|^2.$
 We consider only the
case of zero tree level cosmological constant, i.e., we impose $<G^AG_A>=3$ 
and the order parameter for the supergravity breaking is 
provided by the gra\-vi\-tino mass 
$m_{3/2}^2=e^G.$ A complete scenario of supersymmetry breaking is still
missing. A pragmatic attitude was taken in \cite{brignole1}, where a
parametrization of the supersymmetry breaking was proposed, 
quite independent of its
specific me\-cha\-nism. The fields which participate in the
supergravity breaking were assumed to be the moduli $T_\alpha $ 
and the dilaton $S.$ The parametrization is\footnote{Our definition of 
$\Theta_{\alpha}$ in (\ref{13}) is different from
that in \cite{brignole1}.}

\begin{eqnarray}
G^\beta &=&\sqrt{3} \Theta _\beta  t_\beta  ,  \label{13} \\
  G ^\beta G_\beta &=&3\cos ^2 \theta  ,  \nonumber \\
G^S G_S &=&3\sin ^2\theta .  \nonumber 
\end{eqnarray}
\noindent The angle $\theta $ and the $\Theta _\alpha $'s parametrize the
direction of the goldstino in the $T_\alpha ,S$ space. The normalization 
of the $\Theta _\alpha $ is fixed by (\ref{13}). If (\ref{6}) is assumed we get 
$\sum_{\alpha} n_0^{(\alpha)}
{\Theta_{\alpha}}^2 = \cos ^2 \theta$.
In the presence of the $U(1)_X$ symmetry spontaneously broken
close to the Planck scale there is an additional
contribution to supersymmetry breaking with $<G^\phi G_\phi> \sim$ 
$<\phi>^2.$ 
More generally, any field with a large vev (non negligible compared to $M_P$)
contributes to supersymmetry breaking. 

The soft terms are computed from the usual expressions of supergravity, but
with the flavour non-diagonal K\"{a}hler potential, eq.(\ref{Kahler}). It
is worth noticing that only the lowest power of ${\hat \varepsilon}$ or
$\phi $ have been included in (\ref{Kahler}). Therefore, the predictions 
herebelow have also been derived to the lowest power
of ${\hat \varepsilon}$. In this approximation, it is straightforward
to find
\bea
G^\phi = ( 1-{\sqrt 3} n_\phi^{(\alpha)} \Theta_\alpha ) \phi \ . \label{130}
\eea
Even with $G^\phi \not=0$ in eq. (\ref{130}), 
the parametrization
(\ref{13}) is consistent with the vanishing of the
cosmological constant in the leading order in $\hat \varepsilon$.
 
The soft terms are computed from the scalar potential, which in our case reads ($a=T^{\alpha}, S, \phi,$matter fields). 
\bea
V & = & 
e^{G} 
\left(
 G_A (G^{-1})^{A{\bar B}} G_{\bar B} - 3 
\right) +
{g_X^2 \over 2} 
\left(
K^A \varphi_A \phi_A +  \delta _{GS} M_P^2
\right) ^2 .
\eea 
Since the soft parameters 
are relevant for low-energy phenomenology, it is more appropriate
to express them {\sl after the field redefinition} that brings the kinetic
terms to their canonical forms as consistently done in the following.

Let us first consider the soft scalar masses that have the 
standard supergravity expression \cite{cremmer}
\begin{eqnarray}
\widetilde{m}^2_{i \bj}  & = &
(\widetilde{m}^2_{i \bj})_{F} + 
(\widetilde{m}^2_{i \bj})_{D} 
\nonumber \\
& = &
\left(  G_{i \bj} -
 G^{\alpha} 
{\hat R}_{i \bj \alpha \bbe}
 G^{ \bbe } \right) m_{3/2}^2 +
{g_X} \min (\varphi_i,\varphi_j) G_{i \bj} <D> , \label{150} 
\end{eqnarray}
where ${\hat R}_{i \bj \alpha \bbe}$ is the Riemann tensor of the K\"ahler space  and $<D>$ stands for the 
contribution from the D-term of the $U(1)_X$ gauge group,
\begin{eqnarray} 
D_X & = & {g_X} ( K_{A}\varphi_{A}\phi^{A} +\delta _{GS} M_P^2 ) \nonumber \\
 & = &{g_X} \left( 
-\stackunder{\alpha =1}{\stackrel{p}{\Pi }}
t_\alpha^{n_{\phi}^{(\alpha)}} \phi {\bar{\phi}} +
\min (\varphi_{i},\varphi_{j}) K_{i \bj} {\Phi}^{i} {\bar{\Phi}}^{\bj}
+ \delta _{GS} M_P^2 \right) \ . \label{dsugra}
\end{eqnarray}
In the simple case with
only one singlet field $\phi $, expanding the scalar potential in powers
of $\phi$ we get
\bea
V(\phi)= {\widetilde m}_\phi^2 |\phi|^2 + {1 \over 2} D^2  \ , \label{15}
\eea
where $\widetilde{m}^2_\phi= \left( 1+3 n_{\phi }^{(\alpha )}\Theta
_\alpha ^2 \right)m_{3/2}^2$ is the F-term soft mass of $\phi$
induced by dilaton/moduli breaking. 
In
(\ref{15}) we used the fact that $\phi $ appears in the superpotential 
only through the Yukawa couplings to matter.
The minimization of the potential (\ref{15}) gives  
$g_X <D_X> = \stackunder{\alpha =1}{\stackrel{p}{\Pi }}
t_\alpha^{-n_\phi^{(\alpha )}}
\widetilde{m}^2_\phi$.
The D-breaking is induced by the soft mass $\widetilde{m}^2_\phi$ which is generated by the F-type dilaton/moduli breaking. The breaking of $U(1)_X$
yields a massive real scalar field of mass 
$\sqrt 2 g_X \varepsilon M_P $.  

In the following, we place ourselves in the case where 
\bea
\partial_{\alpha}
Z_{i\bj},\partial_{\bar \alpha} Z_{i\bj} , \partial_{\alpha}
Y_{ij} \simeq 0 \ , \label{151}
\eea
but still they have nontrivial associated modular weights.
For the Yukawa couplings, this happens for example for functions of the
form (keeping only an overall moduli field $T$) $Y \sim c + e^{-T}$ in the
large radius (moduli) limit (see \cite{brignole1} for a more detailed 
discussion on this point). The K\"ahler off-diagonal terms $Z_{i\bj}$ are
related to Yukawas through (\ref{91}) and the explicit moduli dependence
is probably closely related to that of the Yukawas.
 
From (\ref{12}), (\ref{13}) and (\ref{150}), one obtains the expression
for the soft scalar mass matrices as follows,
\begin{eqnarray}
{\widetilde{m}_{i\bj }^2 \over m_{3/2}^2 } = \left( 1+
\varphi_i  + 3 \Theta_\alpha ^2 (n_i^{(\alpha )}+\varphi_i
n_\phi^{(\alpha)}) \right) \delta _{ij} +
\left[ - {1 \over 2}
\left| \varphi _i-\varphi_j\right| + {3 \over 2}(n_{i\bj}^{(\alpha)} 
+ {\bar n}_{i\bj}^{(\alpha)})  \right. \nonumber \\
 \left. \times \Theta_\alpha^2 -\sqrt 3 \Theta_\alpha
(n_{i\bj}^{(\alpha)} \theta_{ji} (\varphi_j - \varphi_i)+ 
{\bar n}_{i\bj}^{(\alpha)} \theta_{ij} (\varphi_i - \varphi_j)) -
3 \Theta_\alpha \Theta_\beta n_{i\bj}^{(\alpha)} {\bar n}_{i\bj}^{(\alpha)}
\right]  {\hat Z}^{\Phi}_{i\bj} 
 \ ,   \label{16} 
\end{eqnarray}
where ${\hat Z}^{\Phi}_{i\bj}= \prod_{\alpha} t_{\alpha}^{{1 \over 2}
(n_{i\bj}^{(\alpha)} + {\bar n}_{i\bj}^{(\alpha)}} Z^{\Phi}_{i\bj }
{\hat \varepsilon}^ {\left| \varphi _i
-\varphi _j\right| }$ and $\theta_{ij}=\theta (\varphi_i - \varphi_j)$.
This general result simplifies in our case of
physical interest, i.e. with the mass hierarchies given solely by the
$U(1)_X$ symmetry. Using eq. (\ref{92}) we find that the 
combination $n_i^{(\alpha)} + \varphi_i n_{\phi}^{(\alpha)}$ is flavour blind 
and we get
\begin{eqnarray}
\widetilde{m}_{i\bi }^2-\widetilde{m}_{j\bj }^2 &=&(\varphi _i -
\varphi _j )m_{3/2}^2 \ 
\label{18} 
\end{eqnarray}
for $\Phi ^i=Q^i,U^i,D^i,L^i,E^i.$ Therefore the splitting in
the diagonal elements of the sfermion masses is independent of the 
parameters $\Theta_{\alpha}$ and proportional to the charge
differences, which fix also the fermion masses. For example, in the Froggatt-
Nielsen case ($\varphi_i \geq \varphi_j$ for $i \geq j$), we have ${m_i^U \over m_j^U} \sim {\hat \varepsilon}^{q_{ij}
+ u_{ij}}$ and we get the fermion-sfermion mass predictions
($\widetilde{m}_{q_i }^2 \equiv \widetilde{m}_{q_{ii}}^2$, etc.)
\bea
m_{3/2}^2 \ ln {m_i^U \over m_j^U} = (\widetilde{m}_{q_i }^2 -
\widetilde{m}_{q_j }^2 + \widetilde{m}_{u_i }^2 - \widetilde{m}_{u_j }^2)
\ ln {\hat \varepsilon}  \label{180}
\eea
and similar relations for down-quarks and leptons.
Also, using (\ref{18}), we deduce that the higher the $U(1)_X$ charge (i.e. 
the smaller the corresponding Yukawa couplings) the larger the soft scalar 
mass. Hence, in that model the spectrum of the matter field superpartners
has {\it inverted hierarchy} compared to that of the associated fermions.

Furthermore, combining (\ref{8}) and (\ref{16}) and introducing the
tree-level gaugino masses $M={\sqrt 3} \sin \theta m_{3/2}$, we obtain
the relations 
\begin{eqnarray}
\widetilde{m}_{q_i }^2+\widetilde{m}_{u_j }^2 + \widetilde{m}_{h_2}^2 &=&M^2
+ (q_i + u_j+h_2 )m_{3/2}^2 \ . \
\label{185} 
\end{eqnarray}
Similar relations are obtained for d-type squarks and sleptons. Notice 
that the combination of charges in the r.h.s. is precisely the power of 
$\hat \epsilon $ in the effective Yukawa couplings. So, since the top Yukawa
is of $O(1)$, $q_3 + u_3 + h_2 = 0.$

Due to the non-diagonal form of the K\"{a}hler potential, the
scalar mass matrices are not diagonal, in contrast to the usual
computations in the literature. 
Moreover, under a stronger assumption $n_{\phi,i\bj}^
{(\alpha)}= {\bar n}_{\phi,i\bj}^{(\alpha)}=0$, we find that the 
flavour-dependent effects in the off-diagonal terms  do not depend 
on the unknown parameters $\Theta_{\alpha}, n_i^{(\alpha)}$.
Remarkably enough, in this case the contribution to 
$(\widetilde{m}_{i\bj }^2)_{F}$ from supersymmetry breaking along
the $\phi $ direction to (\ref{16}) vanishes in the leading order in
$<\phi>/M_P$.

All these equations for scalar masses are to be understood at
energies of the order $M_P,$ 
and lead to low energy relations after renormalization.

The non-diagonal terms in $K$ and  $W$  affect the trilinear
soft terms $V_{ijk}$, too. The general expression for the trilinear
terms corresponding to the fields with $<G^i>=<G_i>=0$ is 
\cite{cremmer}
\begin{eqnarray}
V_{ijk} = \left[ (G^{A} D_{A} + 3) {{W_{ijk}}\over{W}} \right] 
m_{3/2}^2 \ ,
\label{160} 
\end{eqnarray}
where D stands for the covariant derivative in the K\"{a}hler manifold.
Here we give only the expressions for the most contrained case, i.e. where
eqs.(\ref{92}) and  (\ref{151}) are valid, with $n_{\phi,ij}^{(\alpha)}=0$.
Once again we work with canonical normalization of the scalar fields.
With this convention and in the leading order of the small parameter
$\hat \varepsilon$ the connections in the covariant derivatives in
(\ref{160}) take the simple form 
\begin{eqnarray}
G^{\alpha} \Gamma_{\alpha i}^j = \sqrt{3} n_i^{(\alpha)} \Theta_{\alpha} \delta_i^j
+ {{\sqrt{3}} \over 2} \left| \varphi _i-\varphi _j\right|  n_{\phi}^{(\alpha)} \Theta_{\alpha}
{\hat \varepsilon}^ 
{\left| \varphi _i-\varphi _j\right| } Z^{\Phi}_{i\bj } \ ,  
\nonumber \\
G^{\phi} \Gamma_{\phi i}^j = (1-{\sqrt 3} n_\phi^{(\alpha)} \Theta_{\alpha}) 
 ( \varphi _i-\varphi _j) 
\theta (\varphi_i-\varphi_j)  
{\hat \varepsilon}^ 
{\left| \varphi _i-\varphi _j\right| } Z^{\Phi}_{i\bj }
 \ . \label{161} 
\end{eqnarray}

The final result for the triscalar coefficient $V_{ia}^U$, 
for example, reads 
(with ${\hat Y}^{U} = e^{K/2} \, Y^{U}$ )
\begin{eqnarray}
{1 \over m_{3/2}} {V}_{ia}^U  & = & 
\left[ -\sqrt{3} \sin \theta +\left( q_i +u_a + h_2\right) \right]{\hat Y} _{ia}^U \label{17}  \\
& & 
-{1 \over 2} 
\left( \sum_j | q_i-q_j | Z^{Q}_{i\bj } {\hat Y}_{ja} {\hat \varepsilon}^{
|q_i-q_j|} + \sum_b | u_b-u_a| Z^{U}_{a\bar{b} } 
{\hat Y}_{ib} {\hat \varepsilon}^{|u_b-u_a|} \right) \nonumber 
\end{eqnarray}
and similar expressions hold for $V^D$ and $V^L$ with obvious replacements.
Notice that the matrices $\hat Y$ have hierarchical entries and that the
last line in (\ref{17}) contain terms not directly proportional to the
Yukawa coupling ${\hat Y}_{ia}$, but rotated in the flavour space.  The
last terms in (\ref{17}) come from $G^{\phi }D_{\phi },$
namely, from supersymmetry breaking along the $\phi $ direction.
It is useful to introduce the matrices
\bea
{1 \over m_{3/2}}  ({A_L})_{ij} 
& = & 
(q_i + {h_2 \over 2}) \delta_{ij} - {1 \over 2}
|q_i-q_j| Z^{Q}_{i \bj} {\hat{\varepsilon}}^{|q_i-q_j|} \ , 
\label{170}\\
{1 \over m_{3/2}} ({A_R})_{ba} 
& = & 
(u_a + {h_2 \over 2}) \delta_{ab} - {1 \over 2}
|u_a-u_b| Z^{U}_{a {\bar b}} {\hat{\varepsilon}}^{|u_a-u_b|}
\ .  
\nonumber
\eea
Then
\bea
{V}_{ia}^U = -M {\hat Y}^U +  A_L {\hat Y}^U + {\hat Y}^U  A_R
\ . \label {171}
\eea

This parametrization was already used in \cite{brax} in the context of the
models proposed in \cite{brignole1}.
The simplicity of the results follows from nontrivial cancellations between 
$G^{\alpha}$ and $G^{\phi}$ contributions due to eq.(\ref{92}).

As a particular case, in the absence of the $U(1)_X$ symmetry we recover
the minimal MSSM. Therefore, in this case, imposing appropriate modular  
transformations for the renormalizable Yukawa couplings and under the 
assumption that they do not depend explicitly on moduli
fields, we get {\it family-universal} soft terms related by
\bea
\widetilde{m}_{q}^2 + \widetilde{m}_{u}^2 + \widetilde{m}_{h_2}^2 
\hskip-.2cm & = & \hskip-.2cm 
\widetilde{m}_{q}^2 + \widetilde{m}_{d}^2 + \widetilde{m}_{h_1}^2 \, = \, 
\widetilde{m}_{l}^2 + \widetilde{m}_{e}^2 + \widetilde{m}_{h_1}^2 \, = \, 
M^2 \, ,
\nonumber \\
A^{U} 
\hskip-.2cm & = & \hskip-.2cm
A^{D} \, = \, A^{E} \, = \, - M  \, . 
\eea

These simple relations appeared already in the litterature in different
contexts \cite{brignole1,binetruy3} and can be explained here by
the modular invariance conditions of the Yukawa couplings combined with 
dilaton/moduli breaking. It is interesting to compare our results with 
those obtained in \cite{binetruy3}, where the role of the
horizontal symmetry is played by modular symmetries. The essential 
difference is in the predictions for the soft terms which in \cite{binetruy3} turn out to be flavour blind.


\section{Mass terms in the Higgs sector} \label{sec:Higgs}

In this section we discuss the predictions for the mass parameters of the Higgs sector, the soft masses $m^{2}_1,\, m^{2}_2$, the $\mu$ parameter of MSSM and its associated soft breaking term $B\mu$. 

In order to avoid the usual $\mu$-problem of the MSSM, we assume here that both the
$\mu$ and $B\mu$ terms are effective interactions resulting from the 
K\"{a}hler potential after supersymmetry breaking \cite{GM}. Actually, a $H_1H_2$ term 
in the superpotential is forbidden by the $U(1)_X$ symmetry if $h_1+h_2 <0$
(in the case of only one singlet field considered in this section).
For $h_1+h_2=0$, the absence of the corresponding mass term in low energy
string models is equivalent (after decoupling of heavy modes) to 
the presence of massless Higgs doublets in the effective theory. 
Instead, for $h_1+h_2>0$, an effective $\mu-$term of
$O(\epsilon^{h_1+h_2}M_P)$ would be 
possible, which is inconsistent with the proper breaking of the electroweak 
symmetry. This $\mu-$problem could be solved by further symmetries, 
e.g. modular symmetries (provided that they would allow for 
appropriate terms in the K\"{a}hler potential, of course).
The allowed values of $(h_1+h_2)$ are strongly correlated
to the generation of fermion mass hierarchies in this approach with horizontal
$U(1)_X$ symmetry and Green-Schwarz anomaly cancellation. One derives 
\cite{ibanez94} - \cite{nir95} the relation 
$(h_1+h_2)\ln \epsilon \simeq Tr \ln (Y^D/Y^E)$,
which, after substituting the physical fermion masses, favours the values $(h_1+h_2)=0,\, -1$, with 
the values $(+1)$ and $(-2)$ marginally allowed.

For the purpose of generating the $\mu$ and $B\mu$ terms, we consider here 
two classes of K\"{a}hler potentials for the Higgs doublets.
As a first instance, we just extend the general approach of sections 
\ref{sec:generalites} and \ref{sec:soft terms} to include a $H_1H_2$ term in the 
K\"{a}hler potential, with the $U(1)_X$ symmetry restored by powers of $\phi$ or ${\bar{\phi}}$, as follows:
\bea
K & = &
... +
\stackunder{\alpha =1}{\stackrel{p}{\Pi }}t_\alpha
^{n_1^{(\alpha )}} |H_1|^2 + 
\stackunder{\alpha =1}{\stackrel{p}{\Pi }}t_\alpha
^{n_2^{(\alpha )}} |H_2|^2  
\nonumber \\
 &&
+ Z \, \left( 
\left({\phi \over M_P}\right)^{h_1+h_2}H_1 H_2 \, \theta (h_1+h_2) + 
\right. \nonumber \\
&& 
\left.
\stackunder{\alpha =1}{\stackrel{p}{\Pi }}t_\alpha
^{n_1^{(\alpha )}+n_2^{(\alpha )}} ({{\bar \phi } \over M_P})^{-(h_1+h_2)} 
H_1 H_2 \,\theta (-h_1-h_2)
\right. \nonumber \\
&& 
\left.
\hskip 4.0cm + h.c.
\vphantom{\left({\phi \over M_P}\right)^{h_1+h_2}}
\right)   \, ,
\label{Ki}
\eea

\noindent where $Z$ is of $O(1)$ and the dots stand for terms independent of
$H_1,H_2$. Modular invariance of eq. (\ref{Ki}) demands the following relation
\bea
n_{1}^{(\alpha)} + n_{2}^{(\alpha)} + (h_1+h_2)\, n_{\phi}^{(\alpha)}    
& = & 
0 \ ,
\label{rela}
\eea

\noindent if $Z$ is assumed to be $t_{\alpha}-$independent. In turn,
imposing (\ref{rela}) forbids the presence of the $\mu$-term directly
in the superpotential, even in the case $h_1 + h_2 > 0$. From (\ref{Ki}),
on expects the $\mu$-term  to be of the order $O(\epsilon^{|h_1+h_2|}m_{3/2})$, 
also favouring small values of $(h_1+h_2)$.

With the K\"ahler potential completed as in eqs. (\ref{Ki}) and (\ref{rela}),
we return to the question of the uniqueness of the solution (\ref{phi}) for the
vanishing of the D-term of the $U(1)_X$ group. It reads
\bea
D_X 
& = & 
{g_X} 
\left( -\diff{K}{\phi}\phi + \sum_{i=1,2} h_i \diff{K}{H_i}H_i 
+ ... + \delta_{GS} M_P^2 \right)\, .
\label{Dterm}
\eea

\noindent The vanishing of the 
$SU(3)$$\bigotimes SU(2)$$\bigotimes U(1)$ D-terms requires 
$\diff{K}{\ln H_1}$ $ = $$\diff{K}{\ln H_2}$, hence, for the canonically normalized Higgs
fields, ${H_1}=\pm {H_2}= {v}$. In the absence of any relevant term in
the superpotential, there are continuously degenerate solutions satisfying 
$\delta_{GS} M_P^2 ={\phi}^2-(h_1+h_2) {
{v}}^2$.
This degeneracy is removed by supersymmetry breaking assumed in the dilaton and moduli sector, which yields the scale 
$m_{3/2}\ll\epsilon M_P$ and the soft terms. The resulting scalar potential
along the flat direction can be analysed through an expansion in powers
of $\epsilon$. At leading order, we get (for $h_1 + h_2 \not=0$) 
\bea
V = (2+h_1 + h_2 ) m_{3/2}^2 v^2 + const. \ . 
\eea 
Therefore, for $(h_1+h_2) \geq -2$, the minimum is for
$\widetilde{\phi}=\delta_{GS}^{1/2} M_P$, $v=0$ and $D_X=\widetilde{m}_{\phi}^2/g_X$.
The same conclusion holds for $h_1 + h_2 = 0$, as will be evident from the discussion
below eq. (\ref{condi}).
For $(h_1+h_2)=-2$, the continuous degeneracy persists at this level of 
approximation. For $(h_1+h_2) < -2$, the minimum of $V$ is for
$\phi=0$ and $(h_1 + h_2) v^2 = \delta_{GS} M_P^2$, which is physically
uninteresting.
 
We proceed to calculate the effective lagrangian in the Higgs doublet sector. The scalar terms are
obtained from (\ref{150}) and 
the $\mu$ and $B\mu$ effective  parameters are derived from  
the general supergravity expressions for a fermion supersymmetric mass term
$(M_{1/2})_{ij}$ and analytic scalar soft mass $(M_0^2)_{ij}$ ($i,\, j$ are matter indices) \cite{cremmer}

\bea
(M_{1/2})_{ij} 
& = & 
m_{3/2} \left( D_i G_j + {1 \over 3} G_i G_j \right)
\ ,
\nonumber \\
(M_0^2)_{ij}
& = & 
m_{3/2}^2 \left( G^A D_A +2 \right) D_i G_j \ . 
\label{massesoft}
\eea
We find the following results: 
\bea
\widetilde{m}^{2}_i 
& = & 
\left(
1+h_i + 3 \Theta_{\alpha}^2 
\left( n_i^{(\alpha)} + h_i n_{\phi}^{(\alpha)} \right)
\right) m_{3/2}^2 \ ,  
\nonumber \\
\mu 
& = & 
m_{3/2} 
\left(
1+(h_1+h_2)\theta (-h_1-h_2)
\right)
Z \epsilon^{|h_1+h_2|} \ ,
\nonumber \\
B
& = &
m_{3/2} 
\left(
2+(h_1+h_2)\theta (h_1+h_2)
\right)\, \ . 
\label{resuli}
\eea

The simple prediction $B=2$ for $(h_1+h_2) \leq 0 $ arises from a
cancellation between the geometric and $D_X-$term contributions to 
the $B\mu$ analytic coupling, due to the relation (\ref{rela}). 
The latter is absent for $h_1+h_2 >0$ because of the analyticity of the coupling $H_1H_2(\phi/M_P)^{h_1+h_2}$.
Though the parameters $\widetilde{m}_1^2$ and $\widetilde{m}_2^2$ depend on the unknown quantities $\Theta_{\alpha}^2$, their sum does not
\bea
\widetilde{m}_1^2+\widetilde{m}_2^2 
& = & 
(2+h_1+h_2)m_{3/2}^2 \ .
\eea

In order to decide if these predictions are consistent with the requirements of $SU(2)$$\bigotimes U(1)$ breaking, one has to renormalize the 
parameters down to the Fermi scale. In our models 
the relevant parameters take a very simple form. As discussed in section 
\ref{sec:U(1)}, $Y_{33}^U \sim O(1)$ implies 
$h_2+q_3+u_3=0$, which, in turn, gives
\bea
A_t \,\, = \,\, -M &=& - \sqrt{3} m_{3/2} \sin \theta \ ,
\nonumber \\
\widetilde{m}_{h_2}^2+\widetilde{m}_{U_3}^2 &+& \widetilde{m}_{Q_3}^2 
=  M^2 \ . \label{rg}
\eea

Therefore the low energy parameters in the Higgs sector depend 
only on their initial values (\ref{resuli}), on the top mass, 
relatively well-known, and on the gaugino masses parameter, M. 
One gets the following approximate results:
\bea
(\widetilde{m}_1^2+\widetilde{m}_2^2)_{|M_Z}
& = & 
(\widetilde{m}_1^2+\widetilde{m}_2^2) 
-{{1}\over{2}} (10 \rho -2-\rho^2) M^2 \ ,
\nonumber\\
B_{|M_Z}
& = & 
B -{{1-\rho}\over{2}} M \ ,
\nonumber\\
{\mu^2}_{|M_Z} 
& = &
2 (1-\rho)^{1/2} \mu^2 \ ,
\label{renor}
\eea

\noindent where $\rho=m_t^2/m_{t\, crit}^2$ is the ratio between the physical and the 
infrared fixed-point values of the (running) top mass squared. 
Let us consider the following necessary conditions for 
$SU(2)$$\bigotimes U(1)$ breaking in the MSSM model:
\bea
2 |B\mu|_{|M_Z} 
& \leq &
(\widetilde{m}_1^2+\widetilde{m}_2^2+2\mu^2)_{|M_Z} \ , \nonumber \\
{B^2 \mu^2}_{|M_Z} & \geq & (\widetilde{m}_1^2 + \mu^2) {(\widetilde{m}_2^2 + \mu^2)}_{|M_Z} 
\ .
\label{condi}
\eea

For $(h_1+h_2)=-1$, $\mu=0$ indicating that, in this case, the gauged symmetry 
$U(1)_X$ implies an accidental $U(1)_{PQ}$ symmetry. It is well-known
that this symmetry is inconsistent with phenomenology.

For $(h_1+h_2)=0$, one has $\widetilde{m}_1^2+\widetilde{m}_2^2+2\mu^2- 2 B\mu = 2 {(1-Z)}^2 m_{3/2}^2$ at 
the unification scale. This implies (for $Z \not=1$) that the scalar potential at this scale
has a minimum for $H_1 = \pm H_2 = v = 0$.
The relations (\ref{renor}) are satisfied if  
$|M| < (1-Z) m_{3/2}$.
In this case we predict low values for $\tan \beta$. 

For $(h_1+h_2)=1$, $B=3m_{3/2}$, $\mu= Z \epsilon m_{3/2}$, 
$\widetilde{m}_1^2+\widetilde{m}_2^2 = 3 m_{3/2}^2$. The breaking of 
$SU(2)$$\bigotimes U(1)$ then needs a fine-tuning between $M$ and 
$m_{3/2}$ (to $O(\epsilon^2)$), with $\tan \beta \sim O(1/\epsilon^2).$

The case $(h_1+h_2)=-2$, already plagued by vacuum degeneracy at high energies, leads to $B=2m_{3/2}$, 
$\mu\sim \epsilon^2 m_{3/2}$ 
and $\widetilde{m}_1^2+\widetilde{m}_2^2 = 0$. 
These values are inconsistent with (\ref{condi}) even after renormalization.

Summarizing, in this class of models, only those with $(h_1+h_2)=0$ 
seem to stand up to $SU(2)$$\bigotimes U(1)$ breaking requirements, 
if $M < (1-Z) m_{3/2}$. As discussed in section $7$, in this case the 
gaugino mass can help to suppress FCNC effects.
Radiative corrections will affect to some extent the contraints (\ref{condi}),
but a detailed phenomenological discussion of the Higgs sector is beyond the scope of this paper.

 We now turn to a model which have been found in $(2,2)$ superstring  
compactification \cite{AGNT}. The K\"{a}hler potential of such models is
\bea
K=-\ln \left[ (T+{\bar T})(U+{\bar U})-(H_1+{\bar H_2})({\bar H_1} +H_2) \right]+... \ , 
\label{Kii}
\eea
\noindent where $T,U$ are two moduli fields and the dots stand for terms independent of $H_1,H_2$.
The $U(1)_X$ symmetry of (\ref{Kii}) demands $(h_1+h_2)=0$. The modular transformations associated to T and U now read
\bea
T &\rightarrow & {aT-ib \over icT+d} \ , \nonumber\\ 
H_{1,2} &\rightarrow & {1 \over icT+d} H_{1,2} \ , \nonumber\\
U &\rightarrow & U - {ic \over icT+d}  H_1 H_2 \ . 
\label{modutransfo}
\eea

The soft terms are calculated to be
\bea
\widetilde{m}^{2}_{h_i} 
& = & 
\left(
1+h_i - 3 (\Theta_{T}^2+ \Theta_{U}^2)
+ 3 h_{i} (n_{\phi}^{(T)}\Theta_{T}^2 + n_{\phi}^{(U)}\Theta_{U}^2 ) 
\right) m_{3/2}^2 \ ,
\nonumber \\
\mu 
& = & 
\left(
1 +\sqrt{3} \left( \Theta_T + \Theta_U \right)
\right) m_{3/2} \ ,
\nonumber \\
B\mu
& = &
\left( 
2 + 2\sqrt{3} \left( \Theta_T + \Theta_U \right) + 
6 \Theta_T \Theta_U
\right) m_{3/2}^2 \, .
\eea

\noindent This case is much less predictive than the previous one 
(the soft terms depend on the unknown parameters $\Theta_{T,U}$), 
mainly because of the lack of a relation as (\ref{rela}) between
the moduli weights and the $U(1)_X$ charges .  
As noticed in \cite{brignole1} and \cite{BZ} , the inequality
(\ref{condi}) is saturated at the classical level, corresponding 
to a flat direction in the classical potential, even if the degeneracy between $\widetilde{m}_{1}^2$ and $\widetilde{m}_{2}^2$ 
is removed by the $D_X-$terms. From (\ref{renor}) one deduces that 
in order to fulfill the first condition in (\ref{condi}), 
one needs $ M/m_{3/2}  \sim O({1 \over 10}) $, namely, a moduli 
dominated supersymmetry breaking. Furthermore, from the second condition in 
(\ref{condi}) and the requirement of the proper value for
$M_Z$ we get $\widetilde{m}_2^2 \sim O(-M_Z^2)$ and low $\tan \beta$ values. 
It follows from eq.(\ref{rg}) that $\widetilde{m}_{Q_3}, 
\widetilde{m}_{U_3} \sim O ( M_Z )$. Consequently, in this case we expect light gauginos and light third generation
scalars.

It is possible to arbitrarily extend the model to $(h_1+h_2)>0$ by replacing in (\ref{Kii}) and (\ref{modutransfo}), $H_1H_2 \rightarrow \phi^{h_1+h_2} H_1H_2$.
But there is no theoretical basis for such models anymore.

\section{$U(1)\bigotimes U^{\prime}(1)$ horizontal symmetry} \label{sec:2U1}

It has been demonstrated in ref. \cite{leurer93} that, in models with two 
$U(1)\bigotimes U^{\prime}(1)$ symmetries, further suppression of the FCNC 
effects is possible. 

\noindent Hence, we now extend our study to this class of models.
We introduce two SM singlet fields $\phi_1$ and $\phi_2$.
Their charges can be chosen to be 
$\phi_{1} \, (-1,0)$ and $\phi_{2} \, (0,-1)$ (except for the case of proportional charges).
Then, no superpotential term $W(\phi_1, \phi_2)$ can be written for zero vev's of matter 
fields; the vev's of the fields $\phi_1$, $\phi_2$ are fixed by the Fayet-Iliopoulos term.
The superpotential $W$ and the K\"ahler potential $K$
 (with 
$\ve =\phi_{1}/M_P$ and $\vee =\phi_{2}/M_P$) are:

\bea
W  &=& \sum_{ij}
\left[ \,\,
 Y_{ij}^U\;
\theta_{ q_i,u_j,h_2 } \,
\theta_{ q_i^{\prime},u_j^{\prime },h_2^{\prime } }
\,\ve ^{q_i+u_j+h_2}
\vee ^{q_i^{\prime }+u_j^{\prime }+h_2^{\prime }}
Q^i U^j H_2+\right.
\nonumber \\
&&
\hskip .8cm
Y_{ij}^D\;
\theta_{ q_i,d_j,h_1 } \,
\theta_{ q_i^{\prime },d_j^{\prime },h_1^{^{\prime}} }
\,\ve ^{q_i+d_j+h_1}
\vee ^{q_i^{\prime}+d_j^{\prime }+h_1^{\prime }}
Q^i D^i H_1+  \nonumber \\
&&
\hskip .8cm
\left. Y_{ij}^E\;
\theta_{ \ell _i,e_j,h_1 } \,
\theta_{ \ell _i^{\prime },e_j^{\prime},h_1^{\prime } }
\,\ve^{\ell _i+e_j+h_1}
\vee^{\ell _i^{\prime}+e_j^{\prime }+h_1^{\prime }}
L^iE^jH_1
\,\,\right] ,  \nonumber \\
&&  \nonumber \\
K &=& {K_0}\left( T_\alpha ,\bar{T} ^{\bal} \right) 
- \ln\left( S+\bar{S} \right)  
+ \stackunder{\alpha =1}{\stackrel{p}{\Pi }} 
t_\alpha^{n_{\phi_{1} }^{(\alpha )}} \phi_1 {\bar \phi}_1 
+ \stackunder{\alpha =1}{\stackrel{p}{\Pi }} 
t_\alpha^{n_{\phi_{2} }^{(\alpha )}}  \phi_2 {\bar \phi}_2
\nonumber \\
&& \ \ + \sum_{\Phi ^i=Q^i,U^i,D^i,L^i,E^i} 
K^{\Phi}_{i \bj} {\Phi}^{i} {\bar{\Phi}}^{\bj} ,  \nonumber \\
&&  \nonumber \\
K^{\Phi}_{i \bj} &=&
{\delta}_{i \bj}
\stackunder{\alpha =1}{\stackrel{p}{\Pi }}
t_\alpha^{n_i^{(\alpha )}}
+ {Z}^{\Phi}_{i \bj}
\left[\,\, 
\theta _{ij} \,
\theta _{ij}^{\prime } 
\stackunder{\alpha =1}{\stackrel{p}{\Pi }}
t_\alpha^{n_j^{(\alpha )}}
\ve^{\varphi _i - \varphi _j }
\vee^{\varphi _i^{\prime}-\varphi _j^{\prime }}
\right.  \nonumber \\
& & 
\hskip 3.1cm
+\theta _{ji} \,
\theta _{ij}^{\prime}
\stackunder{\alpha =1}{\stackrel{p}{\Pi }}
t_\alpha^{n_j^{(\alpha )}+n_{\phi_{1}}^{(\alpha)}(\varphi_{j}-\varphi_{i})}
\vbe^{\varphi _j - \varphi _i }
\vee^{\varphi _i^{\prime}-\varphi _j^{\prime}} 
 \nonumber \\
&& 
\hskip 3.1cm
+\theta _{ij} \,
\theta _{ji}^{\prime } 
\stackunder{\alpha =1}{\stackrel{p}{\Pi }}
t_\alpha^{n_i^{(\alpha)} +
n_{\phi_{1}}^{(\alpha)} (\varphi _i - \varphi _j)}
\ve
 ^{\varphi _i - \varphi _j }
\vbee
 ^{\varphi _j^{\prime}-\varphi _i^{\prime }} 
\nonumber \\
&&
\hskip 3.1cm
\left.
+\theta _{ji} \,
\theta _{ji}^{\prime }
\stackunder{\alpha =1}{\stackrel{p}{\Pi }}
t_\alpha^{n_i^{(\alpha)}}
\vbe
 ^{\varphi _j - \varphi _i }
\vbee 
 ^{\varphi _j^{\prime}-\varphi _i^{\prime }}
\,\,\right] 
+ ....  \, ,\label{200} 
\eea

\noindent where the previous notation is used and we define 
$\theta_{(a,b,c)} = \theta (a+b+c)$, 
$\theta_{ij} = \theta (\varphi_i-\varphi_j)$ and 
$\theta_{ij}^{\prime} = \theta 
(\varphi_i^{\prime} -\varphi_j^{\prime} )$ .

The $D-$term contributions to the scalar potential are,
\begin{eqnarray*}
V_{D} & = & 
{g_{1}^{2} \over 2} 
\left(
K_{i} \varphi_{i} \Phi^{i} - K_{\phi_{1}}\phi_{1} + \xi_{1}
\right) ^2 +
{g_{2}^{2} \over 2} 
\left(
K_{i} \varphi_{i}^{\prime} \Phi^{i} - K_{\phi_{2}}\phi_{2} + \xi_{2}
\right) ^2 \, ,
\end{eqnarray*}
where $\xi_{1}, \, \xi_{2}$ are the Fayet-Iliopoulos terms. We can, of course, always define one 
linear combination of the two $U(1)$'s which is anomaly free, but
we prefer to work in a basis where the $\phi_i$ charges are simple. Only the simplest and most 
predictive case  
$n_{\Phi,ij}^{(\alpha)} = 
n_{\Phi,i\bj}^{(\alpha)} = 
{\bar n}_{\Phi,i\bj}^{(\alpha)} = 0  $
for 
$\Phi = Q, \,U, \,D, \,E, \,L$ is considered here.

In analogy with the case of one $U(1)$ symmetry, the relation betwen horizontal charges and modular 
weights reads:

\begin{eqnarray}
n_j^{(\alpha)} - n_i^{(\alpha)} & = & 
n_{\phi_{1}}^{(\alpha)} (\varphi _i - \varphi _j) +
n_{\phi_{2}}^{(\alpha)} (\varphi _i^{\prime} - \varphi _j^{\prime}) .
\label{poids-charges}
\end{eqnarray}

The relations (\ref{poids-charges}) enable us to rewrite the matter field metric as:

\begin{eqnarray}
K^{\Phi}_{i \bj} & = &
\stackunder{\alpha =1}{\stackrel{p}{\Pi }}
t_\alpha^{ ( n_i^{(\alpha )} + n_j^{(\alpha)} ) /2}
\left[ 
\delta_{ij} +
{\hat Z}^{\Phi}_{i \bj}
\right] \, ,
\nonumber
\end{eqnarray}
where 
\begin{eqnarray}
{\hat Z}^{\Phi}_{i \bj} & = & 
Z^{\Phi}_{i \bj} 
\stackunder{\alpha =1}{\stackrel{p}{\Pi }}
t_\alpha^{\frac 1 2 
(\,
n_{\phi_{1}}^{(\alpha)}  |\varphi_{i}-\varphi_{j}| + 
n_{\phi_{2}}^{(\alpha)} |\varphi_{i}^{\prime}-\varphi_{j}^{\prime}| \, )}
\nonumber \\
&&
\hskip 1cm
\left[\,\, 
\theta _{ij} \,
\theta _{ij}^{\prime }
\ve^{{\varphi _i}-{\varphi _j}}
\vee^{\varphi _i^{\prime}-\varphi _j^{\prime }} + 
\theta _{ji} \,
\theta _{ij}^{\prime}
\vbe^{{\varphi _j}-{\varphi _i}}
\vee^{\varphi _i^{\prime}-\varphi _j^{\prime}} + \right.
\nonumber \\
&&
\hskip 1cm
\left.
\,\,\,
\theta _{ij} \,
\theta _{ji}^{\prime }
\ve^{{\varphi _i}-{\varphi _j}}
\vbee^{\varphi _j^{\prime}-\varphi _i^{\prime }} +\theta _{ji} \,
\theta _{ji}^{\prime }
\vbe^{{\varphi _j}-{\varphi _i}}
\vbee^{\varphi _j^{\prime}-\varphi _i^{\prime }} 
\,\,\right] \ .
\nonumber
\end{eqnarray}

\noindent The soft scalar mass matrices are as follows:
\begin{eqnarray}
\widetilde{m}_{i \bj}^2  & = & 
\left(  G_{i \bj} -
 G^{A} 
{\hat R}_{i \bj A {\bar B}}
 G^{ {\bar B} } \right) m_{3/2}^2 +
\nonumber \\
&& 
{g_{1}} \min (\varphi_i,\varphi_j) G_{i \bj} <D_{1}> +
{g_{2}} \min (\varphi_i^{\prime},\varphi_j^{\prime}) G_{i \bj} <D_{2}>
\ , 
\end{eqnarray}
where $A,\, B \, = \, \alpha, \, \phi_{1}, \, \phi_{2}$. In absence of the term $W(\phi_{1}, \phi_{2})$ in the superpotential, one finds by minimization of the scalar potential

\noindent ${g_{i}} 
\stackunder{\alpha =1}{\stackrel{p}{\Pi }}
t_\alpha^{n_{\phi_i}^{(\alpha)}}
<D_{i}> \, = \,   
\widetilde{m}_{\phi_{i}}^2$, where 
$ \widetilde{m}_{\phi_{i}}^2 \, = \, 
(1+3 n_{\phi_{i} }^{(\alpha )}
\Theta_\alpha ^2 ) 
m_{3/2}^2$.
\vskip.2cm

The final expression for the scalar masses in the canonical basis reads:
\begin{eqnarray}
\widetilde{m}_{{\hi}{\hbj}}^2  \, &=& \,  
m_{3/2}^2 
\hskip .0cm
\left( \,\,
\Bigl(
1 + \varphi_{\hi} + \varphi_{\hi}^{\prime} +
3 \Theta_\alpha ^2 
(n_{\hi}^{(\alpha )} + 
\varphi_{\hi} n_{\phi_{1}}^{(\alpha)} +
\varphi_{\hi}^{\prime} n_{\phi_{2}}^{(\alpha)})
\Bigr) \delta_{\hi \hj} 
\right.
\nonumber \\
&&
- \frac 1 2 
\Bigl( 
|\varphi_{\hi}-\varphi_{\hj}| + 
|\varphi_{\hi}^{\prime}-\varphi_{\hj}^{\prime}|
\Bigr)
{\hat Z}^{\Phi}_{{\hi} {\hbj}} 
\nonumber \\
&&
\left.
- \, (\varphi _{\hi} - \varphi _{\hj})
(\varphi _{\hj}^{\prime} - \varphi _{\hi}^{\prime})
\theta_{{\hi}{\hj}}
\theta_{{\hj}{\hi}}^{\prime}
{\hat Z}^{\Phi}_{{\hi} {\hbj}}
- \, (\varphi _{\hj} - \varphi _{\hi})
(\varphi _{\hi}^{\prime} - \varphi _{\hj}^{\prime})
\theta_{{\hj}{\hi}}
\theta_{{\hi}{\hj}}^{\prime}
{\hat Z}^{\Phi}_{{\hi} {\hbj}} 
\,\,\, \right) .
\label{smasses}
\end{eqnarray}
As for the one $U(1)_X$ symmetry case, the only dependence on the unknown parameters 
$\Theta_{\alpha}, \,n_{i}^{(\alpha)}$ is contained in a diagonal flavour independent term. This
is a consequence of the eq. (\ref{poids-charges}) which leads to nontrivial 
cancellations between F-terms and D-terms. 
Now, however, supersymmetry breaking along directions $G^{\phi_i}$ do contribute to the soft masses; 
in eq. (\ref{smasses}) the last two lines come from 
$G^{\phi_1}G^{{\bar{\phi}}_2} {\partial K_{i\bj} \over {\partial \phi_1}
{\partial {\bar{\phi_2}}}} + h.c.$

For the trilinear coefficient ${\hat V}_{{\hi}{\ha}}^U$, we obtain, by using (\ref{poids-charges}):
\begin{eqnarray}
{V}_{{\hi}{\ha}}^U \, = \, m_{3/2} & &
\hskip-.6cm
\left( 
\vphantom{{\hat Y} _{{\hi}{\hc}}^U}
\,\,
\left(
-\sqrt{3} \sin \theta +
\bigl( q_{\hi} +u_{\ha} + h_2 \bigr) + 
\bigl( q_{\hi}^{\prime } +u_{\ha}^{\prime } + h_2^{\prime } \bigr)
\right)
{\hat Y} _{{\hi}{\ha}}^U 
\right.
\nonumber \\
&&
\, - \frac 1 2 
\bigl(
|q_{\hi}-q_{\hk}|+
|q_{\hi}^{\prime}-q_{\hk}^{\prime}|
\bigr)
{\hat Z}^{\Phi}_{{\hi} {\hbk}}
{\hat Y} _{{\hk}{\ha}}^U
\nonumber \\
&&
\left.
\,- \frac 1 2 
\bigl(
|u_{\ha}-u_{\hc}|+
|u_{\ha}^{\prime}-u_{\hc}^{\prime}|
\bigr)
{\hat Z}^{\Phi}_{{\ha} {\hbc}}
{\hat Y} _{{\hi}{\hc}}^U
\,\, \right)
 \, ,
\label{trili}
\end{eqnarray}
where we have defined  ${\hat Y}^{U} = e^{K/2}\, Y^{U}$, as in section 4. 
It is straightforward to check that, using the results (\ref{smasses}) and (\ref{trili}), we get predictions similar to eqs.(\ref{18}) and (\ref{185}) with $\varphi_i \rightarrow 
\varphi_i + \varphi_i^{\prime}$.

 An important consequence of eqs.(\ref{smasses}) and (\ref{trili}) is that the flavour off-diagonal
terms can vanish only if 
$\varphi_i = \varphi_j$ and 
$\varphi_i^{\prime} = \varphi_j^{\prime}$. In this case, also the diagonal terms
are flavour independent. However, the corresponding fermion mass matrix will not have
the required hierarchical structure. Thus, it is impossible to have the sfermion 
mass matrices {\it diagonal and degenerate} and simultaneously to keep the hierarchical structure  of
the corresponding fermion mass matrices. This result can be generalized to an arbitrary 
number of abelian symmetries. 
It is, nevertheless, still possible to have degeneracy between some {\it diagonal entries} in sfermion
mass matrices $\widetilde{m}_i^2 = \widetilde{m}_j^2 $ by choosing models with  $\varphi_i + \varphi^{\prime}_i  = \varphi_j + \varphi^{\prime}_j$.
We shall return to this discussion in the next section.

\section{Phenomenological aspects}

We have proposed a class of supergravity models with horizontal abelian gauge
symmetries and modular invariance in which the hierarchies in the fermion
mass spectrum are entirely due to the $U(1)$ symmetries. They have several
interesting phenomenological aspects which can be grouped as very
general qualitative features and more model dependent results. On the general
side, the most important are :

i) high predictivity for the supersymmetric spectrum;
the sfermion spectrum and the Higgs boson spectrum is strongly correlated
with the fermion masses and mixing angles (in the simplest case of one
$U(1)$ it is entirely determined) and shows the family dependence inverse
to fermions (lighter sfermions correspond to heavier fermions). 

ii) generic presence of flavour mixing effects already at the Planck scale
in the squark mass soft terms and trilinear terms; again, these effects are
strongly correlated with the pattern of fermion masses and not only at the
order of magnitude level (like in models with $U(1)$ symmetries alone)
but with the relative magnitude of different terms fixed by the horizontal
charges.

iii) qualitative consistency with the present experimental contraints, which
is remarkable in view of the rigidity of the models; in particular one can
construct models which give FCNC effects at low energy suppressed strongly
enough to meet the experimental limits. However, at the same time our class
of models typically gives FCNC effects which are stronger than expected from
the universality ansatz at the Planck scale  and with predictible 
dependence
on the up-down, left-right sectors. Thus, this class of models is suggestive
of very rich future phenomenology in the domain of FCNC effects, once the
experimental sensitivity is improved. One should stress that we expect FCNC
effects to be only little below the present limits.

On the more model dependent side, we can distinguish the two cases of one
$U(1)$ and two $U(1)$ symmetries. With one $U(1)$ our results are
particularly definite. The only acceptable $U(1)$ charge assignement for the
Higgs fields is $h_1 + h_2 = 0$. The FCNC effects are
predicted to be large, although still marginally acceptable for certain
charge assignments. They have been discussed in some detail in ref. 
\cite{DPS96}, \cite{KK}) and we do not repeat this discussion here.
One should stress that with the moduli-dominated supersymmetry breaking,
i.e. with small strong interaction renormalization effects in the RG
running from $M_P$ to $M_Z$ (see section $5$) the FCNC effects are indeed
at the border line of the present experimental limits \cite{gabbiani89}
for $(\delta_{MM}^{u,d})_{1,2} = {({\Delta M}^{u,d})_{1,2}^2 \over
M_{av}^2}$ $ \ $ ($M=L,R$) and $\delta^{u,d}_{1,2}= \sqrt{{(\delta_{LL}^{u,d})}_{1,2}
{(\delta_{RR}^{u,d})}_{1,2}}$ (defined in the quark mass diagonal basis; $M_{av}$ is an
averaged squark mass).

It was shown in \cite{leurer93} that the constraints from $K^0 - {\bar
K}^0$ mixing can be satisfied in models with two $U(1)$ symmetries and
two mass scales.
Models with two $U(1)$ symmetries give, of course, more freedom in the
$U(1)$ charge assignements consistent with the pattern of Yukawa
matrices and, in consequence, ease the problem of FCNC effects. 

One
can identify certain qualitative features of such models. One way to suppress
FCNC is to impose some partial degeneracy for the diagonal entries in the
squark masses. One can see on general grounds that it is the diagonal
non-degeneracy in the $U(1)$ basis which is the main source of FCNC
effects.
Let us suppose that the Cabibbo angle $\lambda$ is obtained by diagonalizing $Y^U$ ($Y^D$).
Then in the quark physical basis 
$(\delta^u_{LL})_{12} \biggl((\delta^d_{LL})_{12}\biggr) \sim
\max ({{\widetilde{m}_1^2 - \widetilde{m}_2^2} \over 
M_{av}^2} V_{us} ,
{(\widetilde{m}^Q)^2_{12} \over M_{av}^2 })
\simeq V_{us}$. 
This prediction seems to be valid for any model based on an arbitrary 
number of abelian horizontal symmetries. Too large a 
$D^{0}-{\bar D}^{0}$
mixing is predicted (the $K^0 - {\bar
K}^0$ mixing can be suppressed as in \cite{leurer93} or as in our
example below).
In our class of models, 
this result can be 
avoided by a proper choice of charges. 
Indeed, with $q_1+ q^{\prime}_1 = 
q_2+ q^{\prime}_2$ one has  $\widetilde{m}_1^2 = \widetilde{m}_2^2$ and the problem
disappears.

We need to check if this charge assignment is consistent with
Yukawa matrices.
Writing ${{Y^U_{ia}}\over{Y^U_{33}}} \sim \ve^{q_{13}+u_{a3}}
\vee^{q_{13}^{\prime}+u_{a3}^{\prime}} \Theta (q_i + u_a + h_2) 
\Theta (q_i^{\prime} + u_a^{\prime} + h_2^{\prime})$, etc.
one can check that, after imposing $q_1+q_1^{\prime} = q_2+q_2^{\prime}$, 
we get $V_{us} \sim \left( {\ve\over\vee} \right)^{q_{12}^{\prime}}$. So, we 
necessarily need two mass scales. It is easy to construct an acceptable
explicit model. We choose $\ve = \lambda$, $\vee = \lambda^2$ and the charge 
differences
\bea
q_{13} = 1 , q_{23} = 2 , u_{13} = 5 , u_{23} = 2 , d_{13} = 3 , d_{23} = 0 , 
\nonumber\\
q_{13}^{\prime} = 1 , q_{23}^{\prime} = 0 , u_{13}^{\prime} = 0 , u_{23}^{\prime} = 0 , d_{13}^{\prime} = -1 , d_{23}^{\prime} = 0 . 
\label{charges}
\eea

The quark mass matrices are then of the form
$$ Y^U \sim \left( \matrix{ \lam^8 &\lam^5   & \lam^3 \cr   
\lam^7 & \lam^4 & \lam^2 \cr
\lam^5 & \lam^2 & 1} \right)\,\, ,
Y^D \sim \left( \matrix{ \lam^4 &\lam^3   & \lam^3 \cr   
0 & \lam^2 & \lam^2 \cr
0 & 1 & 1} \right).
\label{quarkmat}
$$

For the squark masses we get
\bea
&&
\hskip2.5cm
{(\widetilde{m}_{LL})}^2 \sim \widetilde{m}^2 \left( \matrix{ { 1} &\lam^3   & \lam^3 \cr   
\lam^3 & {1} & \lam^2 \cr
\lam^3 & \lam^2 & 1} \right)\,\, ,
\nonumber \\
&&{(\widetilde{m}^{u}_{RR})}^2 \sim \widetilde{m}^2 \left( \matrix{ 1 &\lam^3 & \lam^5 \cr   
\lam^3 & 1 & \lam^2 \cr
\lam^5 & \lam^2 & 1} \right) \,\, ,
{(\widetilde{m}^{d}_{RR})}^2 \sim \widetilde{m}^2 \left( \matrix{ 1 &\lam^5 & \lam^3 \cr   
\lam^5 & {1} & {0} \cr
\lam^3 & {0} & {1} } \right).
\nonumber 
\eea

From the assignement $q_1+q_1^{\prime} = q_2+q_2^{\prime}$ we get 
${(\widetilde{m}_{LL})}^2_{11} = 
{(\widetilde{m}_{LL})}^2_{22}$. 
The charge assignement (\ref{charges})
gives also $d_2=d_3$ and $d^{\prime}_2=d^{\prime}_3$. Therefore, 
${(\widetilde{m}^{d}_{RR})}^2_{22} = 
{(\widetilde{m}^{d}_{RR})}^2_{33}$ 
and 
${(\widetilde{m}^{d}_{RR})}^2_{23} = 
{(\widetilde{m}^{d}_{RR})}^2_{32}$. 
The most important FCNC effects remain in
$(\delta^{u,d}_{LL})_{12} \sim \lam^3\, , (\delta^u_{RR})_{12} \sim \lam^3\, ,
(\delta^d_{RR})_{12} \sim \lam^5$ 
and 
$\delta^d_{12} \sim \lam^4$.
These signals are below the present experimental bounds but can be 
tested in the next generation of experiments.

Finally, we address the question of new supersymmetric phases which may be
dangerous for CP violations. On general grounds, the soft mass
matrices for left and right handed squarks are hermitian, i.e. the
diagonal terms are real. In the flavour off-diagonal terms (in the
quark mass diagonal basis) arbitrary phases can be present. However, in
any model with proper suppression of the FCNC effects those off-diagonal
terms are suppressed. Thus, in the case of squark masses the problem of new
phases is automatically solved together with the FCNC problem.

For the trilinear terms (L-R mixing terms in the complete squark mass
matrices) the situation is not so simple. It is still true that the
phases in the flavour off-diagonal terms are typically not dangerous,
for the same reason as in case of soft masses. For instance, for one $U(1)$
and in the quark mass diagonal basis we have
 
\bea
{V^U \over m_{3/2}} 
= - {M \over {m_{3/2}}} Y_d^U + (U_L \, A_L \, U_L^{\dagger} ) Y_d^U + Y_d^U (U_R \, A_R \, U_R^{\dagger} )  
\label{179}
\eea

\noindent and similar expressions for $V^D$ and $V^E$. In (\ref{179}), 
$U_L$
and $U_R$ are unitary matrices which diagonalize the mass matrix
$Y_d^U = U_L Y^U U_R^+ $. The result can be expressed in the form
($i \not= j$)
\bea
{v_2 \over m_{3/2}} {(V^U)}_{ij}= A_{ij} 
{\hat \varepsilon}^{|q_{ij}|} m_j^U + B_{ij} 
{\hat \varepsilon}^{|u_{ij}|} m_i^U + \cdots \ , \label{186}
\eea

\noindent where $m_i^U = (m_u,m_c,m_t)$. The $O(1)$ matrices $A_{ij}$ and 
$B_{ij}$ are easily computed; for example,
in the Froggatt-Nielsen hierarchical case,
using the notation
${(U_L)}_{ij}=c_i \delta_{ij} + d_{ij}{\hat \varepsilon}^{|q_{ij}|}$,
we find $A_{ij}= (q_i - q_j) d_{ji}^* c_j - {1 \over 2} |q_{ij}| c_i c_j^*
Z^{\Phi}_{i \bj}$. The dots in (\ref{186}) denote higher order terms.

However, there is no general principle to protect diagonal $A$-terms against
new phases. They can be dangerous for the electric dipole moment of the
neutron. Actually, the relevant phases are $Im (A_{ii} M^* )$, with
$A_{ii}$
defined in eq. (\ref{3}). Consequently, we get
$Im (A_{ii}^U M^* )= (q_i + u_i + h_2) Im (m_{3/2} M^* )$ and all other
possible phases in $G^{\alpha}$ are irrelevant here.  
\noindent
 
\section{Conclusions}

The main purpose of this paper is to propose a theory of flavour where the 
supersymmetric spectrum is completely determined and experimentally testable.
We study effective superstring models
with abelian horizontal gauge symmetries and modular invariances. It is 
shown that the horizontal
charges and the modular weights have to be correlated if the hierarchy
of fermion masses follows solely from the $U(1)$ symmetries.
In consequence, the soft terms in the supersymmetric spectrum,
including the Higgs boson mass terms, are determined by the quark 
masses and mixing angles. This results in a predictive framework 
for the scalar masses. Indeed, the splittings between the diagonal 
as well as the non diagonal entries in the squark and slepton
soft mass matrices turn out to be independent of the direction of
supersymmetry breaking and of the modular weights associated to the
matter fields.

We consider models with one and two $U(1)$
symmetries. The latter have more freedom in the assignement of the 
$U(1)$ charges and allow for stronger suppression of FCNC effects.
However, as a generic feature, models of our type give FCNC effects 
only little below the present experimental limits, and are suggestive 
of very rich future phenomenology in this domain.

\end{document}